Superconductivity of FeSe$_{0.5}$Te$_{0.5}$ Thin Films Grown by Pulsed Laser Deposition

Yoshinori Imai[1, 3, *], Ryo Tanaka[1, 3], Takanori Akiike[1, 3], Masafumi Hanawa[2, 3], Ichiro Tsukada[2, 3], and Atsutaka Maeda[1, 3]

[1] Department of Basic Science, the University of Tokyo, 3-8-1 Komaba, Meguro-ku, Tokyo 153-8902, Japan

[2] Central Research Institute of Electric Power Industry, 2-6-1 Nagasaka, Yokosuka, Kanagawa 240-0196, Japan

[3] Transformative Research-Project on Iron Pnictides (TRIP), JST, Japan.

Abstract

FeSe$_{0.5}$Te$_{0.5}$ thin films with PbO-type structure are successfully grown on MgO(100) and LaSrAlO$_4$(001) substrates from FeSe$_{0.5}$Te$_{0.5}$ or FeSe$_{0.5}$Te$_{0.75}$ polycrystalline targets by pulsed-laser deposition.  The film deposited on the MgO substrate (film thickness ~ 55 nm) shows superconductivity at 10.6 K (onset) and 9.2 K (zero resistivity).  On the other hand, the film deposited on the LaSrAlO$_4$ substrate (film thickness ~ 250 nm) exhibits superconductivity at 5.4 K (onset) and 2.7 K (zero resistivity).  This suggests the strong influence of substrate materials and/or the *c*-axis length to superconducting properties of FeSe$_{0.5}$Te$_{0.5}$ thin films.

Footnotes:

(*) E-mail address: imai@maeda1.c.u-tokyo.ac.jp

I. Introduction

Since the discovery of the superconductivity in F-doped LaFeAsO,[1] considerable studies on the iron-based superconductors have been done and have attracted much attention. FeSe with superconducting transition temperature, $T_c$, of 8 K[2] is one of the iron-based superconductors. Partial substitution of Te for Se raises $T_c$ of FeSe$_{1-x}$Te$_x$ up to a maximum of 14 K.[3] FeSe and FeSe$_{1-x}$Te$_x$ have the simplest structure, namely the tetragonal PbO-type structure, among the iron-based superconductors. Thus, the study of FeSe and related materials is expected to provide a clue to understand mechanism of superconductivity for the iron-based superconductors. In fact, tetragonal FeSe consists of a simple stacking of FeSe layer that is similar to FsAs layer in the other iron-based superconductors, such as LaFeAs(O,F)[1] and Ba$_{1-x}$K$_x$Fe$_2$As$_2$.[4] Therefore, it is expected that FeSe is electrically quasi two-dimensional as other iron-based superconductors. From the first-principle calculations, it is reported that the shape of Fermi surface of FeSe, FeSe$_{1-x}$Te$_x$, and FeTe is very similar to that of the Fe-As based superconductors and that the Fe-As based superconductors and FeSe share a similar superconducting nature.[5]

In general, single crystals and/or epitaxial thin films are desirable for studying intrinsic properties in materials with a finite anisotropy. For example, measurements such as the complex conductivity in the microwave and Terahertz region is anticipated to provide useful information in terms of mechanism of superconductivity, just as in cuprates.[6,7]   These measurements require high-quality single-crystalline thin films that are prepared on a substrate with a small dielectric constant.   In addition, thin films are also requested for possible technological application like Josephson devices.   Some studies on the film growth of $FeSe_{1-x}Te_x$ have been reported.[8-13]   The growth condition and detailed method, however, have not been optimized yet.   For instance, in the films reported in four papers,[8-11] superconductivity has appeared only in films more than 100 nm in thickness, and/or the electrical resistivity above $T_c$ which is one of barometers of a film quality has not been measured precisely, and/or the films have shown only the onset of superconducting transition without zero resistivity.   The other two papers[12,13] have reported that the films deposited on $SrTiO_3$(100) single crystal with high quality show superconductivity at about 15-17 K, which is higher than $T_c$ of polycrystalline samples.[3]   However, $SrTiO_3$ has a very large dielectric constant, which prevents us from applying the complex conductivity measurements in microwave and Terahertz region, and hopefully we need to use a substrate with a small dielectric constant.

In this paper, we report the Pulsed Laser Deposition (PLD) growth of $FeSe_{1-x}Te_x$ superconductors using substrates applicable to high-frequency measurements; MgO and

LaSrAlO$_4$ (LSAO) single crystals. The *a*-axis length of MgO is 4.21 Å, and is far longer than that of FeSe$_{1-x}$Te$_x$ (*a* = 3.76-3.82 Å),[3] while LaSrAlO$_4$ (*a* = 3.76 Å) has a better lattice matching to FeSe$_{1-x}$Te$_x$. In fact, remarkable differences are found between the data of films on MgO and those on LaSrAlO$_4$.

II. Experimental

Two kinds of polycrystalline pellets with the nominal composition of FeSe$_{0.5}$Te$_{0.5}$ and FeSe$_{0.5}$Te$_{0.75}$ were used as the target. The former was prepared according to the method reported in ref. 14 and the latter was synthesized as follows. First, mixed powders of the starting materials, Fe, Se and Te with the molar ratio of 1 : 0.5 : 0.5, were pelletized and heated in the evacuated quartz tube at 953 K for 12 h. Next, reground powders of the reacted pellets and additional Te with molar ratio of 1 : 0.25 were pelletized again and heated in the evacuated quartz tube at 953 K for 2 h. Thin films were grown under vacuum (~ 10$^{-5}$ Torr) by a pulsed laser deposition method using KrF laser (wavelength: 248 nm, repetition rate, *f*: 1-10 Hz). We use a metal mask to prepare FeSe$_{0.5}$Te$_{0.5}$ films in a six-terminal shape for transport measurements. Substrate temperature, $T_s$, was set between 573 and 673 K. Thicknesses of the grown films are measured by Dektak 6M stylus profiler, and estimated to be 20-350 nm.

Considering a possible application to the transmission terahertz spectroscopy, the film thickness less than 100 nm is necessary. The crystal structure and the orientation of films were characterized by x-ray diffraction (XRD) using Cu $K_\alpha$ radiation at room temperature. Electrical resistivity ($\rho$) measurements were carried out by a four-terminal method from 2 to 300 K with magnetic fields up to 13 T applied perpendicular to the film surface.

III. Results and Discussions

Figure 1 shows the XRD patterns of the films deposited at $T_s$ = 573 K with $f$ = 10 Hz on MgO and LSAO. Only the 00$l$ reflections of a tetragonal PbO-structure are observed, which indicates that the films are preferentially oriented along the $c$-axis. The deposition conditions, film thicknesses, $c$-axis lattice parameters, *etc* of several films are summarized in Table I. The $c$-axis lengths of films deposited at $T_s$ = 673 K, namely films F-I, are smaller than those of the other films that are comparable with that of $FeSe_{0.5}Te_{0.5}$ polycrystalline sample ($c$ = 5.97 Å).[3] Considering the dependence of the $c$-axis length on Te content in $FeSe_{1-x}Te_x$ polycrystalline samples,[3] the Te contents in films F-I are likely to be less than those of the other films. This result suggests that the very short $c$-axis length observed in Films F-I is mainly due to the $T_s$-dependent Te content, regardless of substrate materials.

The temperature dependences of the electrical resistivity of the films listed in Table I are shown in Fig. 2. $T_c^{onset}$, defined as the temperature where the electrical resistivity deviated from the normal state behavior, and $T_c^{zero}$, defined as the temperature where the electrical resistivity becomes zero, are estimated from Fig. 2 and are summarized in Table I. Figure 3 is the dependence of $T_c^{zero}$ on $\rho$ at $T$ = 20 K, $\rho_{20K}$. $T_c^{zero}$ of all films can be scaled by $\rho_{20K}$ regardless of the growth conditions (substrate, $T_s$, $f$, $etc$). The films grown at $T_s$ = 673 K do not show a metallic behavior in the temperature dependence of electrical resistivity and superconductivity above 2 K except film G which shows the onset of superconducting transition at 6.4 K. This is probably because the high substrate temperature enhances the deficiencies of Se and/or Te that have a high vapor pressure. This is supported by the fact that the *c*-axis lengths of these four films are very small. $\rho$ at $T$ = 300 K, $\rho_{300K}$, of the films deposited at $T_s$ = 573 K is as small as that of the single crystal of FeSe$_{0.5}$Te$_{0.5}$,[15] and much smaller than that of the polycrystalline sample of FeSe$_{0.5}$Te$_{0.5}$.[3] This indicates that these films are of high quality. Comparing films A, J, and M with films B, L, and O, $T_c^{zero}$ of the films deposited with $f$ = 10 Hz are higher than those of the films with $f$ = 1 Hz. Thus, we found that $T_s$ = 573 K and $f$ = 10 Hz is the best film growth condition.

With respect to the difference of the substrates, we compare films K and L deposited on MgO with films N and O on LSAO. Despite the difference in thickness, $T_c^{onset}$ and $T_c^{zero}$ of film K are as high as those of film L. Notice that the zero resistance shows up even in films that are less than 100 nm in thickness, in contrast to the results reported in ref. 8. In films N and O, on the other hand, $T_c^{onset}$ and $T_c^{zero}$ increase with the film thickness. Another important feature is that $T_c^{onset}$ and $T_c^{zero}$ of the films deposited on LSAO are much lower than those of films on MgO. At first sight, these results look strange in the two senses: First, the thickness dependence on $T_c$ is different between two substrates. Second, $T_c$'s of the films deposited on MgO are higher than those of the films on LSAO, even though film-to-substrate lattice mismatch is far smaller for LSAO than for MgO. From the scanning electron microscopy/energy dispersive x-ray analysis (SEM/EDX), the Fe/Se/Te ratios are 1.03/0.61/0.36 in film L and 1.04/0.56/0.40 in film O, for instance. Considering a measurement error, there is little difference in the Fe/Se/Te ratio between film L and film O. Therefore, the difference in the chemical composition is irrelevant as the origin of the difference between the films on MgO and LSAO. We consider the reason for these surprising results as follows. First, in case of the films grown on MgO substrate, the lattice mismatch is so large that we cannot expect an epitaxial-strain effect for $FeSe_{0.5}Te_{0.5}$ films. The lattice strain might be relaxed within a few layers from the substrate surface. The resultant films have $T_c$ and the $c$-axis lengths which are similar to those of polycrystalline samples. This probably leads to the appearance of superconductivity with zero resistance even in films less than 100 nm in thickness, and also to no change in $T_c$ with increasing film thickness. On the other hand, for the films on LSAO, it is much more reasonable to expect that an epitaxial strain effect works between $FeSe_{0.5}Te_{0.5}$ film and LSAO. This is supported by the fact that the dependence of $T_c^{zero}$ on the $c$-axis length of the films on MgO is different from that of the films on LSAO, as shown in Fig. 4. Fig. 4 means that the $c$-axis length is not the only parameter that determines $T_c$ unlike $\rho_{20K}$ in Fig. 3. In the iron-based superconductors, the pnictogen(chalcogen)-iron-pnictogen(chalcogen) angle[16] or the pnictogen(chalcogen) heights from the iron plane[17] are proposed as the key parameter for superconductivity. These parameters depend not only on the $c$-axis length but also on the $a$-axis length. Thus, the $a$-axis length should also play an important role for superconductivity. The direct measurement of the $a$-axis length is the most appropriate way to discuss the above mentioned issue, which is now in progress. The difference in the dependence of $T_c^{zero}$ on the $c$-axis length indicates that the films on the different substrates receive the different influence of the epitaxial strain effect. Therefore, a finite compressive strain should be applied to FeSe$_{0.5}$Te$_{0.5}$ films deposited on LSAO which has the smaller *a*-axis length than FeSe$_{0.5}$Te$_{0.5}$.  Thus, we consider that the insufficient superconductivity property of films grown on LSAO is due to the negative influence of the compressive epitaxial strain caused by the substrate.  This is consistent with the fact that $T_c$ depends on the film thickness in the films deposited on LSAO, because the thicker the film is the weaker the influence of epitaxial strain is.  The result suggests that the superconductivity of FeSe$_{0.5}$Te$_{0.5}$ is strongly influenced by the epitaxial strain.  It is interesting to see that $T_c$ is enhanced in films deposited on SrTiO$_3$,[12,13] in which the films is believed to suffer from an expansion.  This also seems to be related to the strong pressure effect on the superconductivity in this system.[14,18]  A further systematic investigation of the delicate epitaxial-strain effect is necessary to find a route to higher $T_c$ in this system, which is the subject of subsequent studies.

Finally, we briefly mention the electrical resistivity in the magnetic field.  The effect of applied magnetic field on the electrical resistivity for film K is shown in Fig. 5. $T_c$ decreases linearly with increasing magnetic field.  The upper critical field, $H_{c2}$, which is defined as a field where the resistance becomes half the value of normal-state resistance, is plotted in the inset of Fig. 5 as a function of temperature near $T_c$.  $H_{c2}$ extrapolated to $T$ = 0 K,[19] namely $H_{c2}(0)$, is estimated to be 62.8 T.  That gives a

Ginzburg-Landau coherence length, $\xi_a \sim 22.9$ Å. This is as small as $\xi_a$ of high $T_c$ cuprate superconductors, for example $YBa_2Cu_3O_{7-x}$,[20] suggesting that interesting physics of nano-scale vortex core[21,22] is expected to show up also in this system. Furthermore, since five bands of Fe $3d$ are related to superconductivity in the FeSe system, which is in striking contrast to cuprates with the single Cu $3d_{x2-y2}$ band, this system might provide many novel features in the physics of the mixed state.

IV. Conclusion

We successfully grow the $FeSe_{0.5}Te_{0.5}$ superconducting films by a PLD method. The film deposited on the MgO(100) substrate with film thickness $\sim 55$ nm shows superconductivity at 10.6 K (onset) and 9.2 K (zero resistivity), and has 22.9 Å of the in-plane coherence length. $T_c$ of the film grown on MgO does not depend on the film thickness. For $FeSe_{0.5}Te_{0.5}$ films on MgO, the lattice mismatch is so large that an epitaxial-strain effect is found not to be important. This results in the appearance of superconductivity even in films less than 100 nm in thickness and also in the independence of $T_c$ on the film thickness. On the other hand, the film deposited on the $LaSrAlO_4$ substrate with the film thickness $\sim 250$ nm exhibits superconductivity at 5.4 K (onset) and 2.7 K (zero resistivity). The result that a worse lattice mismatch gives better superconductivity suggests the presence of a strong epitaxial-strain effect only between $FeSe_{0.5}Te_{0.5}$ and $LaSrAlO_4$, which gives a negative influence for superconductivity, in contrast to films on MgO. This is supported by the definite dependence of $T_c$ on the thickness in films on $LaSrAlO_4$. Therefore, we consider that the superconductivity of $FeSe_{0.5}Te_{0.5}$ films is strongly influenced by the epitaxial strain.

V. Acknowledgement

We would like to thank Profs. Jun-ichi Shimoyama and Kohji Kishio at Department of Applied Chemistry, the University of Tokyo for the special supports in the chemical composition analysis of the films.

Figure 1. X-ray diffraction patterns of films grown at $T_s$ = 573 K with $f$ = 10 Hz on (a)MgO (film L) and (b)LSAO (film O). The asterisks represent peaks resulting from the substrate.

Figure 2. The temperature dependence of the electrical resistivity for the films shown in Table I. The inset shows the electrical resistivity near $T_c$. (a) films that are deposited at $T_s$ = 573 K using FeSe$_{0.5}$Te$_{0.75}$ polycrystalline target, namely Films A-E, (b) films that are deposited at $T_s$ = 673 K using FeSe$_{0.5}$Te$_{0.75}$ polycrystalline target, namely Films F-I, (c) films that are deposited at $T_s$ = 573 K using FeSe$_{0.5}$Te$_{0.5}$ polycrystalline target, namely Films J-O. Closed and Open symbols represent the data of the films deposited on MgO, and on LSAO, respectively.

Figure 3. The dependence of $T_c^{zero}$ on electrical resistivity at 20 K, $\rho_{20K}$, shown in Table I. Closed squares represent the data of films deposited on MgO and open circles represent those of films on LSAO.

Figure 4. The dependence of $T_c^{zero}$ on the $c$-axis length of the films shown in Table I. Closed squares represent the data of films deposited on MgO and open circles represent

those of films on LSAO.

Figure 5. The temperature dependence of the electrical resistivity for film K in applied magnetic field. The upper critical field $H_{c2}$ which is defined as a field where the resistance becomes half the value of normal-state resistance is plotted in the inset as a function of temperature near $T_c$.

Table I. The specifications of the grown films.  A Hyphen in the cells of $T_c^{onset}$ and $T_c^{zero}$ means that a film does not show a metallic behavior in the temperature dependence on the electrical resistivity and superconductivity.  "< 2 K" in the cell of $T_c^{zero}$ represents that a film shows only the onset of superconducting transition without zero resistivity above 2 K.

| | target | $T_s$ (K) | $f$ (Hz) | Substrate | Film thickness (nm) | $c$-axis length (Å) | $T_c^{onset}$ (K) | $T_c^{zero}$ (K) |
|---|---|---|---|---|---|---|---|---|
| A | FeSe$_{0.5}$Te$_{0.75}$ | 573 | 1 | MgO | 29 | 5.83 | 5.0 | 3.7 |
| B | FeSe$_{0.5}$Te$_{0.75}$ | 573 | 10 | MgO | 63 | 5.86 | 6.5 | 5.5 |
| C | FeSe$_{0.5}$Te$_{0.75}$ | 573 | 10 | MgO | 212 | 5.93 | 10.3 | 9.5 |
| D | FeSe$_{0.5}$Te$_{0.75}$ | 573 | 10 | LSAO | 52 | 5.84 | 2.5 | < 2 K |
| E | FeSe$_{0.5}$Te$_{0.75}$ | 573 | 10 | LSAO | 348 | 5.90 | 5.7 | 3.7 |
| F | FeSe$_{0.5}$Te$_{0.75}$ | 673 | 3 | MgO | 80 | 5.79 | - | - |
| G | FeSe$_{0.5}$Te$_{0.75}$ | 673 | 3 | LSAO | 53 | 5.80 | 6.4 | < 2 K |
| H | FeSe$_{0.5}$Te$_{0.75}$ | 673 | 10 | MgO | 65 | 5.77 | - | - |
| I | FeSe$_{0.5}$Te$_{0.75}$ | 673 | 10 | LSAO | 50 | 5.78 | - | - |
| J | FeSe$_{0.5}$Te$_{0.5}$ | 573 | 1 | MgO | 135 | 5.86 | 5.7 | 4.8 |

| K | FeSe$_{0.5}$Te$_{0.5}$ | 573 | 10 | MgO  | 55  | 5.92 | 10.6 | 9.2   |
| L | FeSe$_{0.5}$Te$_{0.5}$ | 573 | 10 | MgO  | 164 | 5.90 | 10.0 | 8.9   |
| M | FeSe$_{0.5}$Te$_{0.5}$ | 573 | 1  | LSAO | 190 | 5.84 | 2.7  | < 2 K |
| N | FeSe$_{0.5}$Te$_{0.5}$ | 573 | 10 | LSAO | 80  | 5.86 | 3.9  | < 2 K |
| O | FeSe$_{0.5}$Te$_{0.5}$ | 573 | 10 | LSAO | 250 | 5.88 | 5.4  | 2.7   |

Figure 1

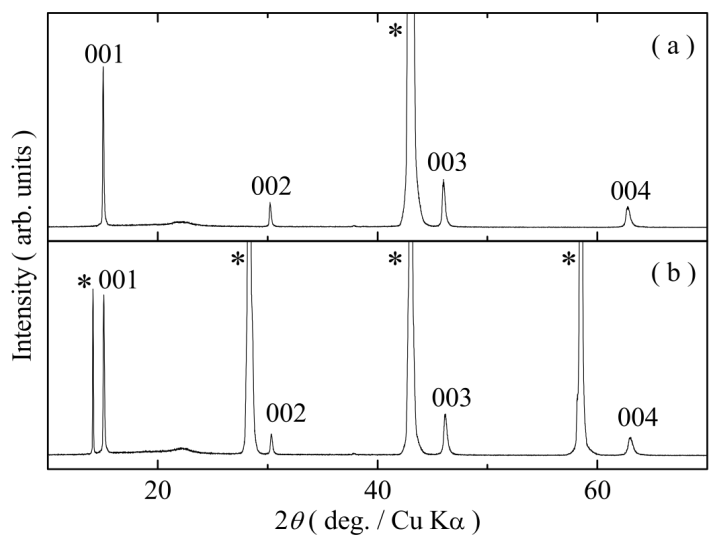

Figure 2

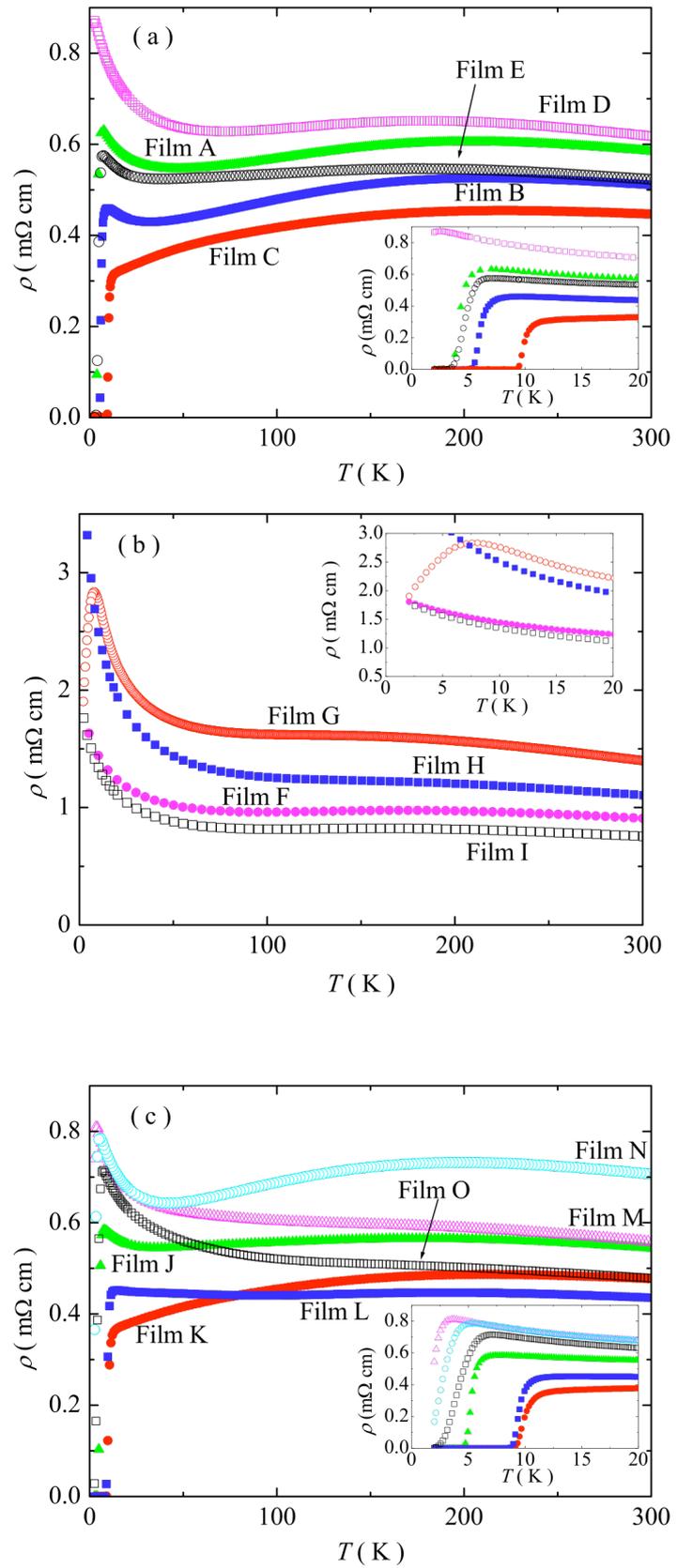

Figure 3

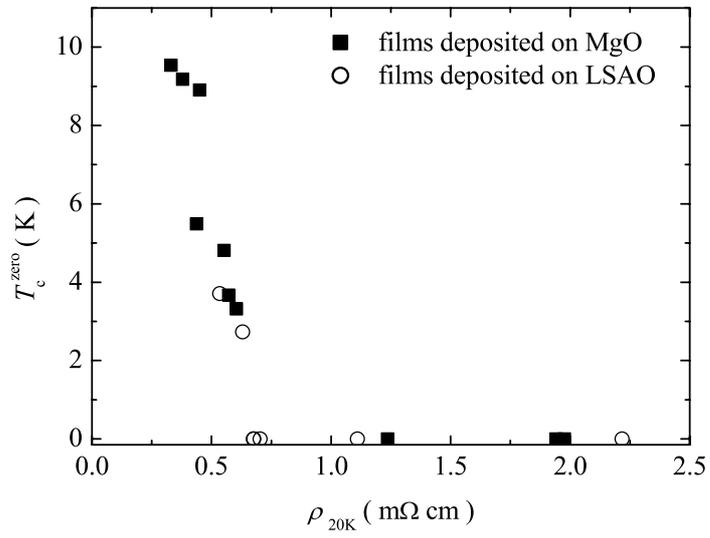

Figure 4

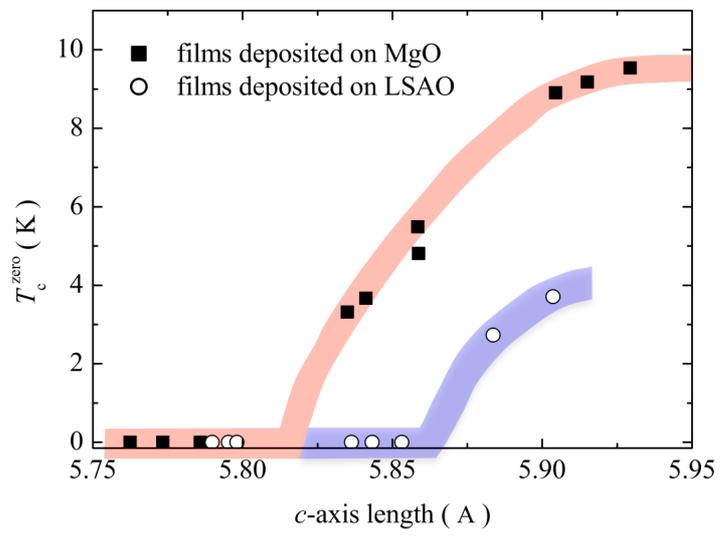

Figure 5

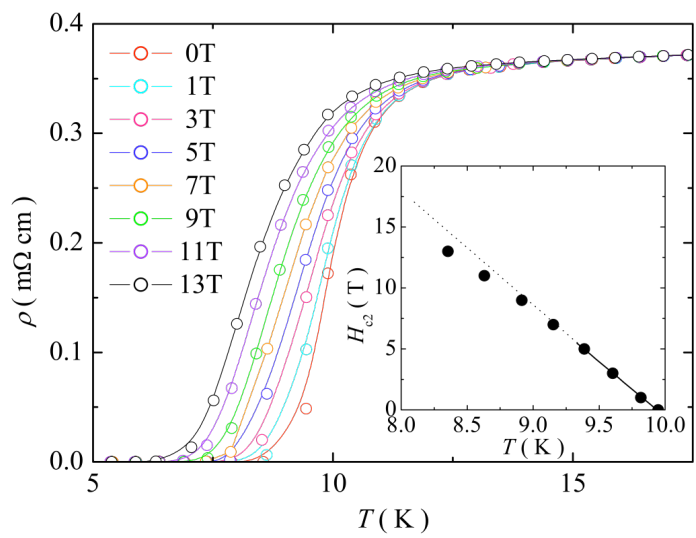